\documentclass[reprint,amsmath,amssymb,superscriptaddress,aps,prl,showkeys]{revtex4-2}
\usepackage[english]{babel}
\usepackage[utf8]{inputenc}
\usepackage{graphicx}
\usepackage{fancyhdr}
\usepackage[active]{srcltx}
\usepackage{setspace}
\usepackage{float}
\usepackage{color}
\usepackage{multirow}
\usepackage{bm}
\usepackage{amsmath, amsfonts, amssymb, mathtools}	
\usepackage{latexsym}
\usepackage{soul}

\usepackage[breaklinks, hidelinks]{hyperref}
\usepackage{xcolor}
\hypersetup{
colorlinks,
linkcolor={red!50!black},
citecolor={blue!50!black},
urlcolor={blue!80!black}
}

\graphicspath{{figures/}}

\newcommand{\etal}{\textit{et al.}}

\begin{document}

\title{Generalized many-body exciton g-factors: magnetic hybridization and non-monotonic Rydberg series in monolayer WSe\textsubscript{2}}

\author{Paulo E. \surname{{Faria~Junior}}}
\thanks{These authors contributed equally}
\email[corresponding authors: ]{paulo@ucf.edu}
\affiliation{Department of Physics, University of Central Florida, Orlando, Florida 32816, USA}
\affiliation{Department of Electrical and Computer Engineering, University of Central Florida, Orlando, Florida 32816, USA}
\affiliation{Institute of Theoretical Physics, University of Regensburg, 93040 Regensburg, Germany}

\author{Daniel \surname{Hernang\'{o}mez-P\'{e}rez}}
\thanks{These authors contributed equally}
\email[corresponding author: ]{d.hernangomez@nanogune.eu}
\affiliation{CIC nanoGUNE BRTA, Tolosa Hiribidea 76, 20018 San Sebasti\'an, Spain}

\author{Tomer \surname{Amit}}
\thanks{These authors contributed equally.}
\affiliation{Department of Molecular Chemistry and Materials Science, Weizmann Institute of Science, Rehovot 7610001, Israel}

\author{Jaroslav \surname{Fabian}}
\affiliation{Institute of Theoretical Physics, University of Regensburg, 93040 Regensburg, Germany}

\author{Sivan \surname{Refaely-Abramson}}
\affiliation{Department of Molecular Chemistry and Materials Science, Weizmann Institute of Science, Rehovot 7610001, Israel}


\begin{abstract}
Magneto-optics of low dimensional semiconductors, such as monolayer transition metal dichalcogenides, offers a vast playground for exploring complex quantum phenomena.
However, current \textit{ab initio} approaches fail to capture important experimental observations related to brightening of excitonic levels and their g-factor dependence.
Here, we develop a robust and general first principles framework for many-body exciton g-factors by incorporating off-diagonal terms for the spin and orbital angular momenta of single-particle bands and many-body states for magnetic fields pointing in arbitrary spatial directions.
We implement our framework using many-body perturbation theory via the GW-Bethe-Salpeter equation (BSE) and supplement our analysis with robust symmetry-based models, establishing a fruitful synergy  between many-body GW-BSE and group theory.
Focusing on the archetypal monolayer WSe\textsubscript{2}, we accurately reproduce the known results of the low-energy excitons including the Zeeman splitting and the dark/grey exciton brightening. 
Furthermore, our theory naturally reveals fundamental physical mechanisms of magnetic-field hybridization of higher-energy excitons (s- and p-like) and resolves the long-standing puzzle of the experimentally measured non-monotonic Rydberg series (1s--4s) of exciton g-factors.
Our framework offers a comprehensive approach to investigate, rationalize, and predict the non-trivial interplay between magnetic fields, angular momenta, and many-body exciton physics in van der Waals systems.

\end{abstract}


\maketitle


\textit{Introduction.} Semiconducting transition metal dichal-cogenides (TMDCs) are a family of two-dimensional, atomically thin layered van der Waals materials \cite{Wang2012, Geim2013, CastroNeto2016, Liu2016, Manzeli2017NRM} that display unique electronic and optical properties, making them promising candidates for ultrathin optoelectronic, photovoltaic, and valleytronic applications \cite{Radisavljevic2011, Andras2013, Hersam2014, Pospischil2014, Withers2015NatMat, Schaibley2016, Ye2017, Mak2016, Mak2018}. 
Particularly, optical properties in TMDC systems are dominated by strongly bound excitons -- quasiparticles resulting from the electron-hole Coulomb interaction \cite{Mak2010PRL, Splendiani2010NL, Qiu2013, Chernikov2014PRL, Urbaszek2018}. 
For monolayers, the combination of broken inversion symmetry and strong spin-orbit interaction imprints selective coupling to circularly polarized light at the inequivalent K and $-$K points in the Brillouin zone\cite{Cao2012, Xiao2012PRL}. As a consequence, direct excitons display valley-dependent optical properties \cite{Sallen2012, Mak2012NatNano}. Moreover, excitons carry intrinsic magnetic moments and effectively couple to external magnetic fields, revealing important effects such as many-body Zeeman shifts and magneto-optical selection rule modifications \cite{Li2014PRL, MacNeill2015, Wang2015TDM, Srivastava2015NatPhys, Aivazian2015NatPhys, Mitioglu2015NL, Stier2016NatComm, Stier2016NL, Schmidt2016PRL, Robert2017PRB, Zhang2017, Arora2018TDM, Stier2018PRL, Liu2019PRB, Chen2019NL, Wang2020NL, Robert2021PRL, Zinkiewicz2021NL, Arora2021JAP, FariaJunior2022NJP, Covre2022Nanoscale, Blundo2022PRL, FariaJunior2023TDM, Pucko2025}.

A fundamental framework for understanding magneto-exciton phenomena is provided by symmetry-based low-energy Hamiltonians \cite{Cho1976PRB, Venghaus1977PRB}. 
Such models explain why optically inactive (dark) excitons brighten under in-plane fields in TMDC monolayers due to spin-flip and valley-mixing terms \cite{Molas2017TDM, Robert2017PRB, Molas2019PRL}.
However, a drawback of such approach is that these models do not offer quantitative insights into the coupling terms.
Conversely, \textit{ab initio} methods based on the many-body GW-Bethe-Salpeter equation (BSE) \cite{Hybertsen1986, Rohlfing1998, Onida1998, Blase2020} explicitly evaluate the underlying Coulomb interactions between the involved quantum states (conduction and valence bands), providing parameter-free information absent in purely symmetry-based models. In particular, the GW-BSE approach captures not only the spatial structure of the single-particle (Bloch) states but also of the excitonic states, achieved by explicitly considering the electron-hole basis and evaluating their many-body interactions.
Within this formalism, the microscopic information about the crystal geometry, atomic nature, and orbital details of the wavefunctions can be incorporated perturbativelly to calculate electronic and excitonic magnetic moments (g-factors) in TMDCs \cite{Chen2019NL, Deilmann2020, Amit2022, FariaJunior2023TDM, Kipczak2023TDM}.
However, existing numerical evaluations of exciton g-factors remain incomplete by neglecting two critical aspects: (1) insights from group-theory analysis and (2) off-diagonal matrix elements (valley, orbital, and spin mixing) in the electronic and excitonic basis. These limitations severely hinder our ability to understand the fundamental aspects, such as spin-valley mixing, decoherence, and relaxation in realistic systems using reliable first principles techniques.

In this paper, we present a robust and general first-principles formalism to describe many-body exciton g-factors by incorporating off-diagonal elements of spin and orbital angular momenta, both in the single-particle and the exciton basis.
These off-diagonal matrix elements are essential for capturing the correct spectral structure in degenerate many-body subspaces and therefore, exciton hybridization under arbitrarily oriented external magnetic fields. 
Our approach synergistically combines many-body perturbation theory within the GW-BSE framework with systematic symmetry-based analysis. 
We validate it by studying the exciton fine structure of monolayer WSe\textsubscript{2}, a prototypical TMDC, reproducing the known results for the exciton Zeeman splitting, as well as the brightening dark/grey excitons in several magnetic field orientations \cite{Robert2017PRB, Molas2017TDM, Molas2019PRL}.
Importantly, we demonstrate that off-diagonal angular momentum terms drive the brightening and hybridization of higher-energy excitons, including s–p mixing, and provide a natural resolution to the long-standing puzzle of the non-monotonic behavior in the excitonic Rydberg series g-factors (1s–4s) \cite{Stier2018PRL, Liu2019PRB, Chen2019NL, Wang2020NL}.
Our formalism opens new opportunities to investigate and predict the role of many-body effects on the non-trivial spin-valley physics of excitons in complex van der Waals systems.


\textit{General theory of exciton g-factors.} 
The Hamiltonian describing two-particle excitations in the presence of an external magnetic field reads $\hat{H} = \hat{H}^\textnormal{BSE} + \hat{\mathbf{g}} \cdot \bm{\mathcal{B}}$ (further details in Sec.~II of the Supplemental Material (SM) \cite{SM}). 
The term $\hat{H}^\textnormal{BSE}$ represents the BSE Hamiltonian, $\hat{\mathbf{g}} = (\hat{g}^x, \hat{g}^y, \hat{g}^z)$ corresponds to the g-factor, $\mathcal{B}_{\epsilon} :=\mu_{B}B_{\epsilon}$, $B_{\epsilon=x,y,z}$ is the external magnetic field, and $\mu_B$ the Bohr magneton. 
In the exciton basis, the BSE Hamiltonian is diagonal, $\langle S |\hat{H}^\textnormal{BSE} |S' \rangle = \Omega_S \delta_{S,S'}$ with $\Omega_S$ being the exciton energies. 
The magnetic coupling  within degenerate and between different exciton subspaces is driven by the external magnetic field and characterized by the Hamiltonian matrix elements $\langle S |\hat{H} |S' \rangle = \sum_\epsilon g^\epsilon_{SS'} \mathcal{B}_\epsilon$, which are excitonic-dressed ``generalized'' g-factor matrix elements
\begin{equation}\label{eq:g-matrix}
g^\epsilon_{SS'}   
= \sum_{vc\mathbf{k}} (\mathcal{A}^S_{vc\mathbf{k}})^\ast \left[ \sum_{c'}  \mathcal{A}^{S'}_{vc'\mathbf{k}} g^\epsilon_{cc'\mathbf{k}} - \sum_{v'} \mathcal{A}^{S'}_{v'c\mathbf{k}} g^\epsilon_{vv'\mathbf{k}} \right].
\end{equation}
Here, $g^\epsilon_{\alpha\alpha'\mathbf{k}} = \langle \alpha \mathbf{k}| \hat{L}^\epsilon + \hat{\Sigma}^\epsilon | \alpha' \mathbf{k} \rangle$ are the electronic g-factor matrix elements \footnote{In this form, the g-factor is simply the total angular momentum, $\mathbf{J} = \mathbf{L} + \mathbf{S}$ with spin-orbit effects incorporated via the basis functions.} accounting for the direct magnetic coupling, $\hat{L}^\epsilon$ ($\hat{\Sigma}^\epsilon$) the components of the orbital (spin) angular momentum operator in the Bloch basis, $|\alpha \mathbf{k}\rangle$, and $\mathcal{A}^S_{vc\mathbf{k}}$ the exciton amplitude obtained from the solution of the BSE.
Eq.~\eqref{eq:g-matrix} extends on previous derivations \cite{Deilmann2020, Amit2022} by considering off-diagonal terms in the exciton g-factor, not only in the excitonic but also in the single-particle electron/hole manifold (band g-factors).
For the latter, we need to consider off-diagonal matrix elements of single-particle operators in the Bloch basis. For instance, the orbital angular momentum in $\hat{\mathbf{z}}$ reads
\begin{eqnarray}\label{eq:L-matrix}
L^z_{\alpha \alpha'\mathbf{k}} = \dfrac{1}{\mathfrak{i} m_0} \Bigg[ \,  \sideset{}{^{\prime}}\sum_{\beta \neq \alpha} \dfrac{p_{\alpha \beta \mathbf{k}}^x p_{\beta\alpha'\mathbf{k}}^y - p_{\alpha\beta \mathbf{k}}^y p_{\beta\alpha'\mathbf{k}}^x}{\epsilon_{\alpha\mathbf{k}} - \epsilon_{\beta\mathbf{k}}} \\ \nonumber 
- \sideset{}{^{\prime}}\sum_{\beta \neq \alpha'} \dfrac{p_{\alpha \beta \mathbf{k}}^y p_{\beta\alpha'\mathbf{k}}^x - p_{\alpha\beta \mathbf{k}}^x p_{\beta\alpha'\mathbf{k}}^y}{\epsilon_{\alpha' \mathbf{k}} - \epsilon_{\beta\mathbf{k}}} \Bigg]
\end{eqnarray}
with $p_{\alpha \beta \mathbf{k}}^{\epsilon} = \langle \alpha \mathbf{k} | \hat{p}^{\epsilon} | \beta \mathbf{k} \rangle$ \footnote{Non-local effects in the band g-factors \cite{SuYing2020, SuYing2021} are not considered. We focus on the combined effect of the excitonic properties within nearly-degenerate subspaces.}, $\mathfrak{i}$ being the imaginary unit, and $m_0$ the bare electron mass. The $\prime$ in the first (second) summation indicates that the $\alpha'$ ($\alpha$) state must be excluded if $\alpha$ and $\alpha'$ are in the same degenerate subset.


\begin{figure}[b]
\begin{center} 
\includegraphics[width=0.95\columnwidth]{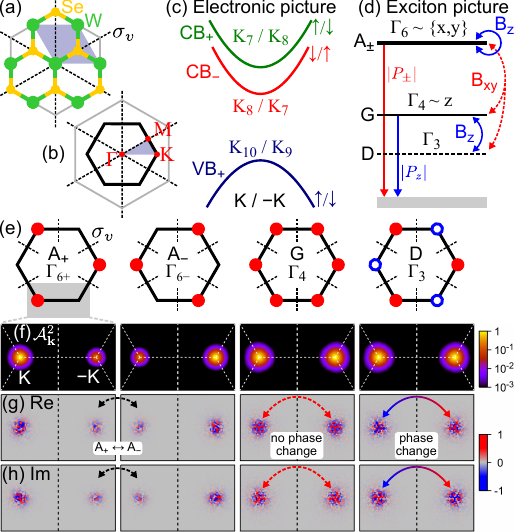}
\caption{(a) Top view of the monolayer WSe\textsubscript{2}. The colored parallelogram indicates the primitive unit cell. (b) First Brillouin zone. The colored triangle indicates the irreducible wedge. The dashed lines in panels~(a,b) indicate the mirror planes $\sigma_v$. (c) Low-energy bands at K$/-$K valleys, including the irreps and spin orientation. (d) Low-energy exciton states at the $\Gamma$-point, identified by their labeling and irreps {(see main text)}. Vertical arrows indicate the allowed optical transitions. Curved arrows indicate couplings via the magnetic field (solid for $\mathbf{B} \parallel z$ and dashed for $\mathbf{B} \parallel x,y$). (e) Schematic representation of the exciton wavefunctions. Closed (open) circles indicate positive (negative) amplitudes. For irrep $\Gamma_6$, the subindex $+(-)$ refers to the exciton wavefunction mainly localized in the $+(-)K$-valley. (f) Absolute, (g) real, and (h) imaginary values of the calculated GW-BSE exciton wavefunctions. The arrows in panels~(g,h) emphasize the effect of $\sigma_v$, i. e., $\Gamma_{6+}(A_+) \leftrightarrow \Gamma_{6-}(A_-)$, no phase change for $\Gamma_4$ (G), and phase change for $\Gamma_3$ (D).}
\label{fig:1}
\end{center}
\end{figure}

\textit{Low-energy excitons.} 
We demonstrate our theoretical approach for the archetypal monolayer WSe\textsubscript{2}, extensively studied via magneto-optics \cite{Aivazian2015NatPhys, Koperski2015NatNano, Robert2017PRB, Molas2017TDM, Stier2018PRL, Liu2019PRB, Chen2019NL, Molas2019PRL, Wang2020NL, Robert2021PRL, Lin2021NatComm}. 
The top view of the TMDC crystals ($D_{3h}$ symmetry group) and the first Brillouin zone are depicted in Figs.~\ref{fig:1}(a,b), respectively. 
We focus on the subspace of low-energy excitons visible in typical magneto-photoluminescence experiments. These excitons arise from the top valence band, VB$_{+}$, and the lowest conduction bands, CB$_{\pm}$, in the vicinity of the K-valleys. 
In Fig.~\ref{fig:1}(c) we show VB$_{+}$ and CB$_{\pm}$, highlighting their irreducible representations (\textit{irreps}) and the direction of the spin expectation value in the out-of-plane direction ($\Sigma^z$) at $\pm$K valleys ($C_{3h}$ point group with complex double group irreps).
The interband optical selection rules at $\pm\text{K}$ can be readily evaluated within group theory (see Sec.~III of the SM \cite{SM}), revealing that the transitions $\text{CB}_{-} \rightarrow \text{VB}_{+} \sim K_{4}$ are optically active with $z$ polarization at $\pm$K points, while $\text{CB}_{+} \rightarrow \text{VB}_{+} \sim K_3 (K_2)$ are optically active with s$_+$ (s$_-$) polarization at K ($-$K) point, in agreement with previous results \cite{Xiao2012PRL, Dery2015PRB, Wang2017PRL}. 
Here, we adopt the group theory nomenclature of Ref.~\onlinecite{koster}.

To evaluate the irreps of the direct (1s-like) excitons, we map the irreps from the K-valleys to the $\Gamma$-point. 
Using the compatibility relations ($C_{3h} \rightarrow D_{3h}$), we find $K_{2} \oplus K_{3} \rightarrow \Gamma_{6}$ and $K_{4} \rightarrow \Gamma_{3} / \Gamma_{4}$, with $\Gamma_{6} \sim x,y$ and $\Gamma_{4} \sim z$. 
The two possibilities of mapping $K_{4}$ to $\Gamma_{3}$ or $\Gamma_{4}$ arise from their distinct behavior under the mirror plane $\sigma_v$ operation, identified in Figs.~\ref{fig:1}(a,b), as a consequence of the intervalley exchange coupling \cite{Glazov2014PRB, Dery2015PRB, Robert2017PRB}.
We employ the typical nomenclature for these excitons \cite{Robert2017PRB, Molas2017TDM, Molas2019PRL}: bright ($\text{A} \sim \Gamma_6$), grey ($\text{G} \sim \Gamma_4$), and dark ($\text{D} \sim \Gamma_3$). 
From the symmetry perspective, A excitons are two-fold degenerate while the D exciton has zero oscillator strength. 
The resulting low-energy subspace, optical selection rules, and coupling to an external \textbf{B}-field are summarized in Fig.~\ref{fig:1}(d), with the schematic representation of the exciton wavefunctions given in Fig.~\ref{fig:1}(e).
 
Our GW-BSE calculations provide the exciton energies and oscillator strengths (Table II in the SM \cite{SM}), supplying a clear identification of these 4 exciton states. 
Notably, the symmetry analysis also allows us to identify the numerical precision of the GW-BSE calculations. 
The energy values (oscillator strengths) fully satisfy the symmetry constraints up to $0.1\,\text{meV}$ ($10^{-3} \; e^2 a^2_0$). 
The exciton energy splitting between G and A excitons is $\sim$52.4 meV while the D-G exciton splitting is $\sim$2.4 meV, in excellent agreement with experimental observations \textcolor{black}{in hBN encapsulated samples} \cite{Wang2017PRL, Robert2017PRB, Molas2017TDM, Molas2019PRL}. 
One relevant aspect of the numerical effects in the GW-BSE calculations is the mixing of the A exciton states, \textit{i.e.}, the two states do not emit light with completely circular polarization. 
This is a common effect in first-principles calculations whenever the irreps are (nearly) numerically degenerate \cite{Cassiano2024SciPost, Zhang2023vasp2kp}.
To verify the features from Fig.~\ref{fig:1}(e) at the GW-BSE level, we display in Figs.~\ref{fig:1}(f-h) the density ($\rho^S_{\mathbf{k}} =\sum_{vc}\left|\mathcal{A}_{vc\mathbf{k}}^{S}\right|^2$), real ($\text{Re}\{\mathcal{A}^S_{\mathbf{k}}\} = \sum_{vc} \text{Re}\{ \mathcal{A}_{vc\mathbf{k}}^{S} \}$), and imaginary ($\text{Im}\{\mathcal{A}^S_{\mathbf{k}}\} = \sum_{vc} \text{Im}\{ \mathcal{A}_{vc\mathbf{k}}^{S} \}$) values of the computed \textit{ab initio} exciton wavefunctions. 
Because of the numerical degeneracy in the A exciton subspace, the wavefunctions are not fully localized in $\pm$K but are still connected by the mirror plane $\sigma_v$. 
For the G and D excitons, the sign change is visible in the real and imaginary parts, evidenced by the dashed (G) and solid (D) arrows.

Knowing the excitons' symmetry allows us to incorporate the effect of external magnetic fields, which transform as pseudovector objects ($B_{x,y} \sim \Gamma_5$ and $B_{z} \sim \Gamma_2$). 
The resulting symmetry-allowed couplings in the low-energy excitons are shown in Fig.~\ref{fig:1}(d), revealing that out-of-plane fields ($\mathbf{B} \parallel z$) yield Zeeman splitting physics for the $A$ exciton subset and mixing of D and G excitons, while in-plane fields ($\mathbf{B} \parallel x,y$) introduce exciton mixing between A and D/G states. We emphasize here the relevance of our general formalism: mixing effects can only be captured by off-diagonal matrix elements, which are also relevant for degenerate subsets such as the A exciton.
The details of the symmetry analysis and connection to the microscopic contributions of electron g-factors in the exciton basis are given in Sec.~III of the SM \cite{SM}. 
We incorporate the magnetic field within GW-BSE by numerically evaluating  Eq.~\eqref{eq:g-matrix}, including the \textbf{k}-space extension of the exciton wavefunctions coupled via the full matrices $\Sigma$ and $L$. 

\begin{figure}[htb]
\begin{center} 
\includegraphics[width=0.95\columnwidth]{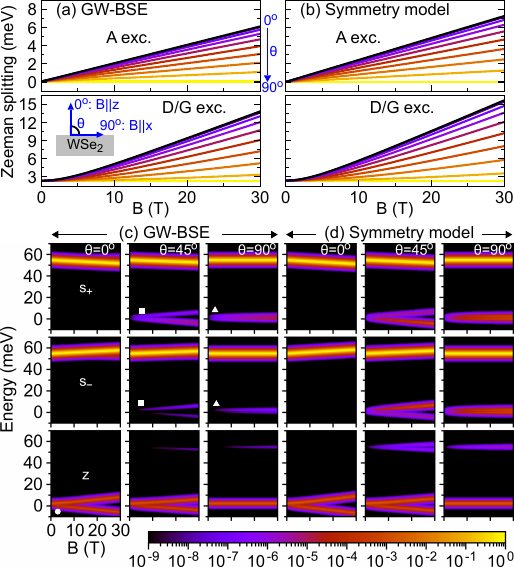}
\caption{Zeeman splitting of the low-energy excitons under applied magnetic field at different angles, $\theta$, for (a) the GW-BSE and (b) the symmetry-adapted model. The top (bottom) panels correspond to the A (D/G) excitons.
Calculated absorption via the (c) GW-BSE and (d) symmetry-adapted model under applied magnetic field oriented at different angles ($\theta = 0^\circ, 45^\circ, 90^\circ$) for s$_+$ (top row),  s$_-$ (middle row), and z (bottom row) polarizations. The spectra are normalized to the maximum value of the s$_+$ emission. For each transition, a broadening was applied using a sech function with 1 meV full-width at half-maximum. The closed circle, squares, and triangles highlight important brightening signatures.}
\label{fig:2}
\end{center}
\end{figure}

To validate our approach, we show in Figs.~\ref{fig:2}(a-b) the Zeeman splitting of A and D/G excitons for different orientations of the magnetic field.
In the symmetry model we assume $\delta$-like exciton wavefunctions fully localized at $\pm$K points, as in previous works \cite{Molas2019PRL}. 
Besides the difference in the Zeeman splitting ($\sim$1.2 meV at 30 T, due to the \textbf{k}-space exciton wavefunction spread \cite{Deilmann2020, Amit2022, FariaJunior2023TDM}), all the dependencies match accordingly, revealing that the exciton couplings are properly captured by our general \textit{ab initio} formalism. 

The magneto-optical signature of exciton mixing is observed in the intensity of the photoluminescence spectra \cite{Robert2017PRB, Molas2017TDM, Molas2019PRL}.
In Figs.~\ref{fig:2}(c-d) we present the calculated absorption spectra in logarithmic scale \footnote{In photoluminescence experiments, the optical spectra incorporate the optical selection rules and the exciton occupation (Boltzmann distribution function \cite{Zhumagulov2020JCP} with rapid exponential decay). Therefore, the emission of A, G, and D excitons display similar intensities \cite{Robert2017PRB, Molas2017TDM, Molas2019PRL, Lin2021NatComm, Dirnberger2021SciAdv}, which can also influenced by the aperture of the objective lens used to collect the emitted light \cite{Lin2021NatComm}.} for different magnetic field orientations ($\theta = 0^\circ, 45^\circ, 90^\circ $). 
Our calculations reproduce all the relevant mixing-induced brightening mechanisms observed experimentally: (1) brightening of the D exciton due to the D-G mixing for out-of-plane fields \cite{Robert2017PRB} (closed circle); (2) brightening of both G and D excitons in tilted magnetic fields \cite{Molas2017TDM} (squares); and (3) brightening of both D and G excitons for in-plane fields \cite{Molas2019PRL} (triangles). 
The mixing-induced exciton brightening under magnetic fields is a direct consequence of the off-diagonal elements of the spin and orbital angular momenta matrices, correctly incorporated in the general expression, Eq.~\eqref{eq:g-matrix}, of the exciton g-factor.

\begin{figure}[htb]
\begin{center} 
\includegraphics[width=0.95\columnwidth]{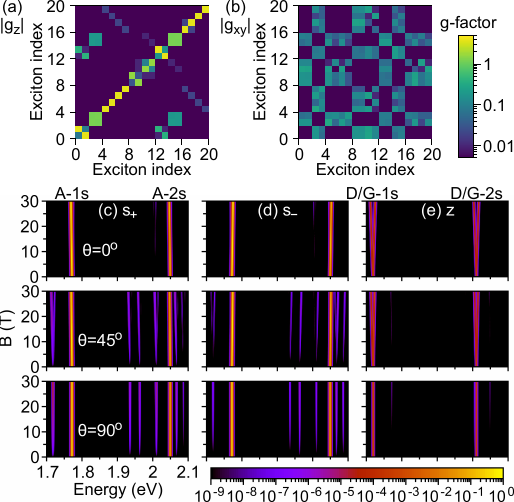}
\caption{Absolute value of the (a) out-of-plane g-factor matrix, $g_z$, and (b) in-plane g-factor matrix, $|g_{xy}| = \sqrt{g_x^2 + g_y^2}$, via Eq.~\eqref{eq:g-matrix}, as a function of the exciton index, demonstrating the significance of the off-diagonal terms.
Calculated absorption for high-energy excitons as a function of the magnetic field oriented at different angles ($\theta = 0^\circ, 45^\circ, 90^\circ$) using the GW-BSE approach for (c) s$_+$, (e)  s$_-$, and (e) z polarizations.
The values are normalized to the maximum value of the s$_+$ emission. We apply the same broadening as in Fig.~\ref{fig:2}. 
Several exciton peaks emerge around 1.95 eV, visible for $\theta = 45^\circ, 90^\circ$. 
We consider the nonzero oscillator strengths only for s-like exciton states.}
\label{fig:3}
\end{center}
\end{figure}

\textit{Magnetic-hybridization of high-energy excitons.} While low-energy excitons are easily described by effective models, high-energy excitons (often called \textit{excited} excitons) pose a greater challenge due to the increasingly denser excitonic manifold of available states with enhanced intervalley exchange and spin-orbital mixing.
To emphasize the significance of the exciton hybridization at higher energies, we present in Fig.~\ref{fig:3}(a-b) the absolute values of the out-of-plane and in-plane g-factors for the lowest 20 excitons(see Sec.~V of the SM \cite{SM} for a larger exciton subset).
Notably, both g-factor components display significant presence of off-diagonal elements responsible for hybridizing states that belong to different exciton subspaces, completely absent in previous studies \cite{Deilmann2020, Amit2022}. 
Moreover, g-factors of p-like excitons (indices 5-12 and 17-20) have similar magnitude of their s-like counterparts (indices 1-4 and 12-16). In Sec.~V of the SM \cite{SM} we present the exciton wavefunctions.

In Fig.~\ref{fig:3}(c-e) we display the calculated absorption for s$_\pm$ and z polarization for the exciton subset shown in Fig.~\ref{fig:3}(a-b). 
For $s_{\pm}$ polarizations, external magnetic fields lead to the brightening of several optically inactive excitonic states. 
In particular, we reveal a clear signature of the mixing of s-p excitons, recently observed by external in-plane electric fields in monolayer WSe\textsubscript{2} \cite{Zhu2023PRL}. For z-polarized light, we recover the brightening of the excited D/G states, see Fig.~\ref{fig:2}(c). 
Similar to the recently proposed s–p mixing of excitons in van der Waals heterostructures \cite{Cao2025}, this brightening cannot be captured by purely (Wannier or symmetry-based) effective models, underscoring the importance of including the full complexity of the excitonic spectrum using the generalized GW-BSE treatment.

\begin{figure}[htb]
\begin{center} 
\includegraphics[width=0.95\columnwidth]{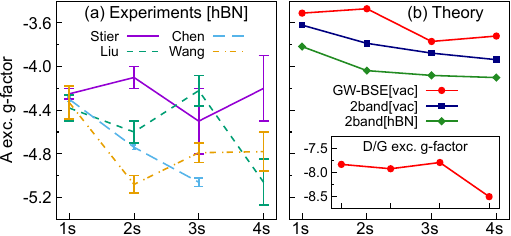}
\caption{(a) Experimental g-factors of the Rydberg series (1s--4s) of the A exciton by Stier \etal~\cite{Stier2018PRL}, Liu \etal~\cite{Liu2019PRB}, Chen \etal~\cite{Chen2019NL}, and Wang \etal~\cite{Wang2020NL}.
(a) Calculated g-factors via the GW-BSE approach (circles), using an effective two-band model (squares), and via the two-band model considering hBN encapsulation (diamonds).  Notably, the observed non-monotonic trends can only be accurately reproduced using our \textit{ab initio} GW-BSE generalized g-factor theory. Inset: g-factors of D/G Rydberg excitons, also exhibiting non-monotonic features.}
\label{fig:4}
\end{center}
\end{figure}

\textit{Rydberg series of g-factors.} 
To showcase the robust capabilities of our approach, we address the long-standing puzzle and conflicting experimental reports of the Rydberg series of the A exciton g-factors for the so-called 1s--4s states. Fig.~\ref{fig:4}(a) compiles the available experimental g-factors in hBN-encapsulated monolayer WSe\textsubscript{2} from Refs. \cite{Stier2018PRL, Liu2019PRB, Chen2019NL, Wang2020NL}. These experiments reveal an overall decreasing trend together with clear non-monotonic signatures, varying slightly due to environmental and sample-dependent factors \cite{Mennel2018NatComm, Raja2019NatNano, Kolesnichenko2020TDM}.

In Fig.~\ref{fig:4}(b), we present our calculated g-factors via the full \textit{ab initio} GW-BSE approach, revealing that non-monotonic features naturally emerge from our formalism (see Sec.~V of the SM \cite{SM} the wavefunctions of 1s--4s states).
Notably, the g-factors for D/G Rydberg excitons also exhibit non-monotonic dependencies, see inset of Fig.~\ref{fig:4}(b).
We also evaluate the g-factors using an effective (parabolic) two-band model incorporating the dielectric screening of vacuum and hBN-encapsulation (see Sec.~I of the SM \cite{SM}).
The effective model systematically yields a monotonic behavior and fails to reproduce the oscillations obtained in the GW-BSE g-factor calculations via Eq.~ \eqref{eq:g-matrix}. 
These non-monotonic features cannot be explained by simplified models but naturally emerge from our generalized GW-BSE formalism, which incorporates orbital and spin mixing induced by external magnetic fields. 
Our results firmly indicate that the interplay between the excitonic fine structure and magnetic response is highly nontrivial and strongly dependent on the exciton state, emphasizing the need for first-principles-based approaches when interpreting high-resolution magneto-optical measurements.

\textit{Conclusions.} In this paper, we developed a robust and general framework based on \textit{ab initio} many-body GW-BSE formalism that incorporates the interplay of the spin and orbital angular momenta via the exciton g-factors. 
This approach takes into account the hybridization of single-particle bands and many-body states through off-diagonal matrix elements of spin and orbital angular momenta.
Moreover, we establish a previously-unexplored synergy between GW-BSE and group theory models of excitonic subspaces.
We validate our approach for the archetypal TMDC monolayer WSe\textsubscript{2}, capturing and rationalizing the observed results of the exciton Zeeman splitting as well as brightening of the optically dark/grey excitons by in-plane/tilted magnetic fields.
We also explore the brightening of high-energy excitons, a challenging task for pure symmetry-based models, emphasizing that many-body off-diagonal components of the g-factor are crucial to capture the magnetic mixing of exciton states.
These results imply that interpreting high-energy features in magneto-optics requires caution, as nominally distinct excitonic subspaces can strongly hybridize. 
Furthermore, the robustness of our approach allows us to unveil the non-monotonic behavior of the Rydberg series of exciton g-factors, a long-standing puzzle observed by several experimental groups \cite{Stier2018PRL, Liu2019PRB, Chen2019NL, Wang2020NL}.
Our novel approach provides a robust foundation to study many-body effects of multidirectional magneto-excitons in other complex two-dimensional materials and van der Waals heterostructures. It is particularly relevant to study non-trivial spin-valley dynamics \cite{Raiber2022NatComm} and unusual topological and chiral excitons \cite{Hou2023PRL, Xie2024PNAS} as well as many-body effects in the emergent field of orbitronics \cite{Rappoport2025}.


\begin{acknowledgments}

The authors acknowledge the financial support of the Deutsche Forschungsgemeinschaft (DFG, German Research Foundation) SFB 1277 (Project No. 314695032) and the European Research Council (ERC) Starting Grant No. 101041159.
P.E.F.J. acknowledges the computational resources of the Advanced Research Computing Center of the University of Central Florida.
D. H.-P. acknowledges the funding from the Diputaci\'on Foral de Gipuzkoa through Grants 2023-FELL-000002-01, 2024-FELL-000009-01, from the Spanish {MICIU/AEI/10.13039/501100011033} and FEDER, UE through Project No. PID2023-147324NA-I00, and the computational resources of the Max Planck Computing and Data Facility (MPCDF) cluster.
T.A. acknowledges support from the Azrieli Graduate Fellows Program. 

\end{acknowledgments}


\bibliography{biblio}

\end{document}


\title{Generalized many-body exciton g-factors: magnetic hybridization and non-monotonic Rydberg series in monolayer WSe\textsubscript{2}}

\author{Paulo E. \surname{{Faria~Junior}}}
\thanks{These authors contributed equally}
\email[corresponding author: ]{paulo@ucf.edu}
\affiliation{Department of Physics, University of Central Florida, Orlando, Florida 32816, USA}
\affiliation{Department of Electrical and Computer Engineering, University of Central Florida, Orlando, Florida 32816, USA}
\affiliation{Institute of Theoretical Physics, University of Regensburg, 93040 Regensburg, Germany}

\author{Daniel \surname{Hernang\'{o}mez-P\'{e}rez}}
\thanks{These authors contributed equally}
\email[corresponding author: ]{d.hernangomez@nanogune.eu}
\affiliation{CIC nanoGUNE BRTA, Tolosa Hiribidea 76, 20018 San Sebasti\'an, Spain}

\author{Tomer \surname{Amit}}
\thanks{These authors contributed equally.}
\affiliation{Department of Molecular Chemistry and Materials Science, Weizmann Institute of Science, Rehovot 7610001, Israel}

\author{Jaroslav \surname{Fabian}}
\affiliation{Institute of Theoretical Physics, University of Regensburg, 93040 Regensburg, Germany}

\author{Sivan \surname{Refaely-Abramson}}
\affiliation{Department of Molecular Chemistry and Materials Science, Weizmann Institute of Science, Rehovot 7610001, Israel}

\maketitle

\tableofcontents


\clearpage

\section{Computational details}

\subparagraph{Geometries.}
For our calculations, we considered the geometry of monolayer WSe\textsubscript{2} with an in-plane lattice parameter fixed to the experimental value of 3.289~\AA~\cite{Schutte1987, Kronik2023} with a Se-Se distance of 3.277~\AA. A vacuum separation of 15.9\,\AA\ was considered in the out-of-plane direction to minimize interactions between periodic replicas. 

\subparagraph{Density functional theory}
$\,$

\textit{Quantum Espresso}. 
We performed density functional theory (DFT) calculations using the \textsc{Quantum Espresso} package \cite{Giannozzi2009, Giannozzi2017, Giannozzi2020}.
For the exchange-correlation functional, we employed the non-empirical PBE generalized gradient approximation (GGA) \cite{Perdew1996}.
\textsc{Quantum Espresso} utilizes a plane-wave basis set, for which we set a kinetic energy cut-off of 90~Ry.
Spin-orbit interaction was included by employing fully relativistic optimized norm-conserving Vanderbilt pseudopotentials from the PseudoDojo library \cite{Dojo2019}.
The self-consistent charge density was converged on a $24 \times 24 \times 1$ \textbf{k}-point grid.

\textit{WIEN2k.} To validate and cross-check DFT calculations performed within \textsc{Quantum Espresso}, we also performed DFT calculations using the all-electron full-potential implementation of WIEN2k\cite{wien2k}, one of the most accurate DFT codes available\cite{Lejaeghere2016Science}. We also employed the PBE exchange-correlation functional, a \textbf{k}-grid of $30 \times 30 
\times 1$, and a self-consistent convergence criteria of 10$^{-6}$ $e$ for the charge and 10$^{-6}$ Ry for the energy. The core and valence electrons are separated by $-6$ Ry. We used orbital quantum numbers up to 10 within the atomic spheres and the plane-wave cutoff multiplied by the smallest radius of the muffin-tin sphere is set to 8. The radius of the muffin-tin sphere considered are 2.41 bohr for W and 2.30 bohr for S. For the inclusion of SOC, core electrons are considered fully relativistically and valence electrons are treated within a second variational step\cite{Singh2006}, with the scalar-relativistic wave functions calculated in an energy window from $-10$ to $8$ Ry. 

We find that the resulting DFT gap for the monolayer case at the K point is 1263.0 (1283.4) meV obtained with WIEN2k (\textsc{Quantum Espresso}).

\subparagraph{GW.} We compute the quasi-particle energy spectrum within the GW approximation,  using the \textsc{BerkeleyGW} package\cite{Deslippe2012, Rohlfing1998, Rohlfing2000, Wu2020}. As a starting point, we use DFT energies and wavefunctions computed with \textsc{Quantum Espresso}, then apply a one-shot non-self-consistent GW calculation (G\textsubscript{0}W\textsubscript{0})\cite{Hybertsen1986}, employed using the spinor implementation within the \textsc{BerkeleyGW} package. This implementation includes non-perturbative spin-orbit coupling corrections\cite{Deslippe2012, Wu2020, Louie2022}. 
The dielectric function was computed using the generalized plasmon-pole model of Hybertsen and Louie \cite{Hybertsen1986}.
We used a $6 \times 6 \times 1$ uniform \textbf{q}-grid, including a total of 4000 empty and occupied states, and applied a 25~Ry cut-off for the dielectric function.
To accelerate convergence with respect to \textbf{k}-point sampling, we refined the Brillouin zone sampling near $\mathbf{q}= \mathbf{0}$ by incorporating 10 additional \textbf{q}-points within a non-uniform neck subsampling scheme \cite{Qiu2017}.
The Coulomb interaction was truncated along the out-of-plane direction to prevent spurious coupling between periodic replicas \cite{Ismail2006}.  
We find a GW gap at the K point for the monolayer of 2311.63 meV, which is 1028.23 meV larger than the DFT gap.

\subparagraph{BSE.} The Bethe-Salpeter equation (BSE) was solved using the Tamm-Dancoff approximation (see also Sec. \ref{app:bse}) within the \textsc{BerkeleyGW} package\cite{Deslippe2012, Rohlfing1998, Rohlfing2000, Wu2020}.
The electron-hole interaction kernel matrix elements of the Bethe-Salpeter Hamiltonian were computed on a uniform Monkhorst-Pack $24 \times 24 \times 1$ coarse \textbf{k}-grid and subsequently interpolated to a $90 \times 90 \times 1$ uniform fine \textbf{k}-grid, which was subsequently employed for the absorption and g-factor calculations. 
An energy cut-off of 5~Ry was applied to the dielectric matrix in the BSE electron-hole kernel matrix elements.
The electron-hole kernel was computed including 24 bands (12 valence and 12 conduction).
For the absorption calculations, a reduced set of 12 bands (6 valence and 6 conduction) was considered.  

\subparagraph{Orbital angular momentum matrix elements.}
Single-particle angular momentum matrix elements were calculated extending previous work by the some of the authors\cite{Wozniak2020PRB, Amit2022, FariaJunior2022NJP}, see also Sec. \ref{app:angular_momentum}. 
In the \textsc{Quantum Espresso} calculations, we included a summation over 600 occupied and empty bands and checked the convergence of the results with respect to the total number of bands. 
For the calculation of the orbital angular momentum matrix elements, we used DFT energies and performed a scissor-shift of the conduction bands using the GW bandgap at the K point. 
In WIEN2k calculations, the energy window ($-10$ to $8$ Ry) yields a total of 60 occupied and 1182 empty bands, enough to converge the orbital angular momentum calculations\cite{FariaJunior2022NJP, FariaJunior2023TDM}.

\subparagraph{Effective two-band model for excitons.} The g-factors obtained within an effective two-band model follow the approach discussed in Refs. \cite{Kipczak2023TDM, FariaJunior2023TDM} and do not incorporate off-diagonal terms neither in the single-particle nor in the exciton basis. 
Specifically, we treat conduction and valence bands with a parabolic dispersion with effective masses given by $0.2650$ and $-0.3519$, respectively, obtained from the WIEN2k calculations. 
The exciton wavefunctions are calculated considering the Rytova-Keldysh potential \cite{Rytova1967, Keldysh1979} with an effective monolayer thickness of 6.55 \textrm{\AA} and a dielectric constant of 15.6 \cite{Laturia2018npj}. 
The dielectric constant of the surrounding dielectric environment is taken as $\varepsilon = 1$ to mimic air and $\varepsilon = 5$ to mimic hBN. The $k$-dependence of the $X$ exciton g-factor has a convex dispersion with minimum at the $K$ valley (consistent with previous calculations\cite{Deilmann2020, FariaJunior2023TDM, Gobato2022NL, Kipczak2023TDM}) and can be approximated by a parabola for $g_{X}(k)=g_{X}^{(0)}+g_{X}^{(2)}k^{2}$ with coefficients $g_{X}^{(0)} = -4.12$ and $g_{X}^{(2)} = 96.84$ (these values include the scissor shift in the band gap, see also Table\ref{tab:DFTslg}). 
Within this approach, the exciton g-factor, i.e., the single-particle parabolic dependence $g_{X}(k)$ weighted by the exciton wavefunction $F(k)$ from the two-band model (see also Eq.~\ref{eq:gzdiag}), we found $g_{X} = 2\pi\int\text{d}k\,k\,g_{X}(k)\left|F(k)\right|^{2} = g_{X}^{(0)}+2\pi g_{X}^{(2)}\int\text{d}k\,k^{3}\left|F(k)\right|^{2}$. 
The numerical grid used for the exciton energies and wavefunctions considers a circular region with radius 0.5 \textrm{\AA} extracted from a 181$\times$181 square grid from -0.5 \textrm{\AA} to 0.5 \textrm{\AA}, leading to 25445 effective k-points. 
The binding energies for the 1s, 2s, 3s, and 4s states are 430.2 (131.2) meV, 186.5 (25.7) meV, 109.5 (10.2) meV, and 71.8 (5.8) meV, respectively, for $\varepsilon = 1$ ($\varepsilon = 5$).


\clearpage

\newpage

\section{Operators representation in the excitonic basis}

\subsection{Preliminaries: basis set}\label{app:bse}
In this work, we consider the ensemble of excitons $\{S\}$ with energies $\{\Omega_S\}$, which are obtained by numerically solving the BSE within the Tamm-Dancoff approximation\cite{Rohlfing2000} for a finite number of conduction and valence quasi-particle bands.
This equation is an effective eigenvalue problem $\hat{H}^\textnormal{BSE}|S \rangle = \Omega_S |S \rangle$, where the matrix elements of the Hamiltonian written in the electron-hole basis read
\begin{equation}
    \hat{H}^\textnormal{BSE}_{cv\mathbf{k};c'v'\mathbf{k}'} = (E_{c\mathbf{k}} - E_{v\mathbf{k}}) \delta_{c,c'}\delta_{v,v'}\delta_{\mathbf{k},\mathbf{k}'} + K^{\textnormal{eh}}_{vc\mathbf{k}; v'c'\mathbf{k}'}.
\end{equation}
Here, $E_{c\mathbf{k}}$ (resp. $E_{v\mathbf{k}}$) are the quasi-particle energies of the conduction (resp. valence) bands, $K^{\textnormal{eh}}_{vc\mathbf{k}; v'c'\mathbf{k}'}  = \langle vc \mathbf{k}|\hat{K}^{\textnormal{eh}}|v'c' \mathbf{k}' \rangle$ are the matrix elements of the  electron-hole interaction kernel.

It is understood that by employing this equation, each exciton state is expressed as the coherent superposition of electron-hole pairs 
\begin{equation}\label{e0}
    |S \rangle = \sum_{vc\mathbf{k}} \mathcal{A}^S_{vc\mathbf{k}} |v \mathbf{k}  \rangle |c\mathbf{k} \rangle,
\end{equation}
where $\mathcal{A}^S_{vc\mathbf{k}}$ is the exciton amplitude and $|v \mathbf{k} \rangle |c\mathbf{k} \rangle $ is a short-hand notation for $|v \mathbf{k} \rangle \otimes |c\mathbf{k} \rangle $. In other words, exciton states are spanned in the conduction-valence (or electron-hole) basis, which belongs to the tensor product space $\mathcal{H}_\textnormal{ex} = \mathcal{H}_{\textnormal{e}} \otimes \mathcal{H}_{\textnormal{h}}$.
For simplicity in this work, we only consider the case of direct excitons, assuming negligible coupling to phonons or electrons and a vanishingly small momentum transfer coming from the absorbed light.  

\subsection{Angular momentum operators in the excitonic basis}\label{app:angular_momentum}
We start our derivations by considering the exciton spin operators. 
The exciton spin operator directed along the spatial direction $\hat{\bm{\epsilon}}$, with $\epsilon = x,y,z$, is defined as
\begin{equation}\label{eq:spin_decomp}
    \hat{\Sigma}^\epsilon = \hat{\Sigma}^\epsilon_e + \hat{\Sigma}^\epsilon_h := \hat{\Sigma}_e^\epsilon \otimes 1\!\!1_h - 1\!\!1_e \otimes \hat{\Sigma}_h^\epsilon,
\end{equation}
where the minus sign is related to the definition of the hole operator and we promoted the single-particle operators in the electron and hole subspaces to combined electron-hole Hilbert space by using a tensor product with the $2\times 2$ identity operator, $1\!\!1$, in each respective subspace. 
Each spin operator is a matrix, with its elements given in the Bloch basis, $|\alpha \mathbf{k} \rangle$, by
\begin{equation}\label{eq:spin_bloch}
\Sigma^\epsilon_{\alpha\alpha'\mathbf{k}} = \langle \alpha \mathbf{k}| \sigma^\epsilon |  \alpha' \mathbf{k} \rangle,
\end{equation}
and measured in units of $\hbar/2$. Here, $\sigma^\epsilon$ correspond to the Pauli matrices, with the following representation
\begin{equation}
\sigma^x = 
\begin{pmatrix}
0 & 1 \\
1 & 0
\end{pmatrix}, \qquad
\sigma^y = 
\begin{pmatrix}
0 & -i \\
i & 0
\end{pmatrix},  \qquad
\sigma^z = 
\begin{pmatrix}
1 & 0 \\
0 & -1
\end{pmatrix}.
\end{equation}

Using Eq. \ref{eq:spin_decomp}, we can explicitly compute matrix elements of the operator in the excitonic basis to find
\begin{equation}\label{e1}
\langle S | \hat{\Sigma}^\epsilon | {S'} \rangle = \sum_{vc\mathbf{k}} \Bigg[ \sum_{c'} (\mathcal{A}^S_{vc\mathbf{k}})^\ast \mathcal{A}^{S'}_{vc'\mathbf{k}} \Sigma^\epsilon_{cc'\mathbf{k}} - \sum_{v'} (\mathcal{A}^S_{vc\mathbf{k}})^\ast \mathcal{A}^{S'}_{v'c\mathbf{k}} \Sigma^\epsilon_{vv'\mathbf{k}} \Bigg].
\end{equation}
In the absence of degenerate excitonic subspaces, only the diagonal components of Eq. \eqref{e1}, $\langle \hat{\Sigma}^\epsilon \rangle_S :=  \langle S | \hat{\Sigma}^\epsilon | {S} \rangle$ would be necessary, thus, Eq. \eqref{e1} simplifies to
\begin{equation}
\langle \hat{\Sigma}^\epsilon \rangle_S = \sum_{vc\mathbf{k}} \Bigg[ \sum_{c'} (\mathcal{A}^S_{vc\mathbf{k}})^\ast \mathcal{A}^{S}_{vc'\mathbf{k}} \Sigma^\epsilon_{cc'\mathbf{k}}  - \sum_{v'} (\mathcal{A}^S_{vc\mathbf{k}})^\ast \mathcal{A}^{S}_{v'c\mathbf{k}} \Sigma^\epsilon_{vv'\mathbf{k}} \Bigg],
\end{equation}
Furthermore, if the spin is a good quantum number at the single-particle level, so that all the transitions are also spin-conserving and the single-particle spin operator is diagonal in the Bloch basis, we find the simplified form 
\begin{equation}
\langle \hat{\Sigma}^\epsilon \rangle_S  = \sum_{vc\mathbf{k}} |\mathcal{A}^S_{vc\mathbf{k}}|^2 (\Sigma^\epsilon_{c\mathbf{k}} - \Sigma^\epsilon_{v\mathbf{k}}) ,
\end{equation}
where $\Sigma^\epsilon_{\alpha\mathbf{k}}:= \Sigma^\epsilon_{\alpha\alpha \mathbf{k}}$. This expression was already suggested in Ref. \onlinecite{Ruan2022} when considering the expectation value of the spin of non-degenerate excitons in the $\hat{\mathbf{z}}$ direction in Bi/SiC. We also note that similar approaches to the one described here have recently been used to investigate exciton effects on the shift current. \cite{Esteve-Paredes2025}

The orbital angular momentum can be decomposed in a analogous way to Eq. \eqref{eq:spin_decomp}. For the single-particle states, we follow Ref. \onlinecite{Wozniak2020PRB}, and the matrix elements in the Bloch basis of the (symmetrized) angular momentum operator (here we choose for simplicity the $\hat{\mathbf{z}}$ direction, but the matrix elements in the $\hat{\mathbf{x}}$ and $\hat{\mathbf{y}}$ directions are obtained by a straightforward permutation of indices) can be written as 
\begin{equation}\label{L1}
L^z_{\alpha \alpha'\mathbf{k}} = \dfrac{1}{\mathfrak{i} m_0} \Bigg[ \, \sideset{}{^{\prime}}\sum_{\beta \neq \alpha} \dfrac{p_{\alpha \beta \mathbf{k}}^x p_{\beta\alpha'\mathbf{k}}^y - p_{\alpha\beta \mathbf{k}}^y p_{\beta\alpha'\mathbf{k}}^x}{\epsilon_{\alpha\mathbf{k}} - \epsilon_{\beta\mathbf{k}}} 
- \sideset{}{^{\prime}}\sum_{\beta \neq \alpha'} \dfrac{p_{\alpha \beta \mathbf{k}}^y p_{\beta\alpha'\mathbf{k}}^x - p_{\alpha\beta \mathbf{k}}^x p_{\beta\alpha'\mathbf{k}}^y}{\epsilon_{\alpha' \mathbf{k}} - \epsilon_{\beta\mathbf{k}}} \Bigg],
\end{equation}
where we employed units of $\mathcal{B}_z = \mu_B B_z$. Here, the Bohr's magneton is given by $\mu_B = |e| \hbar/(2m_0)$, $B_z$ is the external applied magnetic field along $\hat{\mathbf{z}}$,  $p^\epsilon_{\alpha\beta\mathbf{k}}$ corresponds to the matrix elements of the canonical momentum in the Bloch basis and $\epsilon_{\alpha\mathbf{k}}$ are the Kohn-Sham (or GW) energies. The $\prime$ symbol in the first (second) summation indicates that the $\alpha'$ ($\alpha$) state must be excluded if $\alpha$ and $\alpha'$ are in the same degenerate subset, leading to the equation
\begin{equation}\label{L2}
L^z_{\alpha \alpha'\mathbf{k}} = \dfrac{1}{\mathfrak{i} m_o} \sum_{\beta \neq \alpha,\alpha'} \Bigg[ \dfrac{p_{\alpha \beta \mathbf{k}}^x p_{\beta\alpha'\mathbf{k}}^y - p_{\alpha\beta \mathbf{k}}^y p_{\beta\alpha'\mathbf{k}}^x}{\epsilon_{\alpha\mathbf{k}} - \epsilon_{\beta\mathbf{k}}} - 
 \dfrac{p_{\alpha \beta \mathbf{k}}^y p_{\beta\alpha'\mathbf{k}}^x - p_{\alpha\beta \mathbf{k}}^x p_{\beta\alpha'\mathbf{k}}^y}{\epsilon_{\alpha' \mathbf{k}} - \epsilon_{\beta\mathbf{k}}} \Bigg].
\end{equation}
We note that expressions \eqref{L1} and \eqref{L2} naturally takes into account degenerate single-particle states and can be derived within the context of the \textbf{k.p} perturbation theory or by employing the relation between the position operator and the Berry connection \cite{Yafet1963g, Bhowal2021PRB, Urru2025}. 

Following the same steps as for the exciton spin, the excitonic matrix elements of the angular momentum operator in the $\hat{\bm{\epsilon}}$ direction simply read
\begin{equation}
\langle S | \hat{L}^\epsilon | {S'} \rangle = \sum_{vc\mathbf{k}} \Bigg[ \sum_{c'} (\mathcal{A}^S_{vc\mathbf{k}})^\ast \mathcal{A}^{S'}_{vc'\mathbf{k}} L^\epsilon_{cc'\mathbf{k}} \notag  - \sum_{v'} (\mathcal{A}^S_{vc\mathbf{k}})^\ast \mathcal{A}^{S'}_{v'c\mathbf{k}} L^\epsilon_{vv'\mathbf{k}} \Bigg],
\label{eq:L_exc}
\end{equation}
with $L^\epsilon_{\alpha\alpha'\mathbf{k}} = \langle \alpha \mathbf{k}| \hat{L}^\epsilon |  \alpha' \mathbf{k} \rangle$.

\subsection{Generalized excitonic g-factor}

We now define the operator $\hat{\mathbf{g}} := \hat{\mathbf{L}} + \hat{\bm{\Sigma}}$, where we considered $g_0 \simeq 2$, so that each component of this vector operator is expressed in reduced units of $\mathcal{B}_\epsilon := \mu_B B_\epsilon$.
Straightforwardly inherited from its angular momentum composition, and as shown in Sec. \ref{app:angular_momentum}, the exciton g-factor is in general not  diagonal in the excitonic basis. We can easily obtain the general matrix elements once the angular momentum matrix elements are known,
\begin{align}
\langle S | \hat{g}^\epsilon | {S'} \rangle  = \sum_{vc\mathbf{k}} \Bigg[ \sum_{c'} (\mathcal{A}^S_{vc\mathbf{k}})^\ast \mathcal{A}^{S'}_{vc'\mathbf{k}} g^\epsilon_{cc'\mathbf{k}} - \sum_{v'} (\mathcal{A}^S_{vc\mathbf{k}})^\ast \mathcal{A}^{S'}_{v'c\mathbf{k}} g^\epsilon_{vv'\mathbf{k}} \Bigg],
\end{align}
with $g^\epsilon_{\alpha\alpha'\mathbf{k}} = L^\epsilon_{\alpha \alpha'\mathbf{k}} + \Sigma^\epsilon_{\alpha \alpha' \mathbf{k}}$.

Importantly, this expression for the exciton g-factor generalizes the one employed in previous works\cite{Deilmann2020, Amit2022}, where degenerate exciton subspaces were not taken into consideration and thus only the diagonal matrix elements, as well as only the out-of-plane $\hat{\mathbf{z}}$ direction, $\langle \hat{g}^z \rangle_S := \langle S |\hat{g}^z | S \rangle$, was considered, 
\begin{equation}
\langle \hat{g}^z \rangle_S = \sum_{vc\mathbf{k}} |\mathcal{A}_{vc\mathbf{k}}^S|^2 (g^z_{c\mathbf{k}} - g^z_{v\mathbf{k}}),
\label{eq:gzdiag}
\end{equation}
where $g^z_{\alpha\mathbf{k}} = L^z_{\alpha \mathbf{k}} + \Sigma^z_{\alpha \mathbf{k}}$.


\clearpage

\newpage

\section{Symmetry analysis}\label{app:mono}

 The group theory nomenclature used here follows Ref.~\onlinecite{koster}.

\subsection{Zero magnetic field}

The dipole matrix element mediated by the linear momentum operator between conduction ($c$) and valence ($v$) bands is given as $\mathbf{p}^\epsilon_{cv} = \left\langle c,\pm \text{K} | \mathbf{p} \cdot \hat{\epsilon} | v,\pm \text{K} \right\rangle$. Within group theory $\mathbf{p}^\epsilon_{cv} = \Gamma^*_{c} \oplus \Gamma_{\epsilon} \oplus \Gamma_{v}$, with $\Gamma_{c(v)}$ being the irreps of conduction (valence) bands and $\Gamma_{\epsilon}$ is the irrep of the vector coordinate $\epsilon = x,y,z$. The transition $\text{CB}_{-} \rightarrow \text{VB}_{+}$ at K ($-$K) yields $K_{8}^{*} \otimes K_{10} = K_{4}$ ($K_{7}^{*} \otimes K_{9} = K_{4}$), while $\text{CB}_{+} \rightarrow  \text{VB}_{+}$ at K ($-$K) yields $K_{7}^{*} \otimes K_{10} = K_{3}$ ($K_{8}^{*} \otimes K_{9} = K_{2}$). The irreps related to the in-plane circular polarization (s$_\pm$) are $K_{2} \sim x-iy, \; K_{3} \sim x+iy$ and for out-of-plane linear polarization ($z$), the irrep is $K_{4} \sim z$.

\subsection{Introducing the magnetic field}

In transition metal dichalcogenide (TMDC) monolayers, the-low energy excitons with zero-momentum (located at the $\Gamma$ point) have the following irreducible representations (irreps) within the $D_{3h}$ symmetry group:
\begin{align}
\text{Bright}(A) & \sim\Gamma_{6} \sim \{x,y\} \rightarrow 2\text{ dimensional},\nonumber \\
\text{Grey}(G) & \sim\Gamma_{4} \sim z \rightarrow 1\text{ dimensional},\nonumber \\
\text{Dark}(D) & \sim\Gamma_{3}\rightarrow 1\text{ dimensional}.
\label{eq:exc_irreps}
\end{align}

Based on symmetry considerations, we can derive the Hamiltonian for the Zeeman splitting for these low energy states. Specifically, the Hamiltonian term that couples to the magnetic field reads:
\begin{equation}
\hat{H}_{\text{Z}}(\mathbf{B})=\hat{H}_{x}\hat{\mathbf{x}}+\hat{H}_{y}\hat{\mathbf{y}}+\hat{H}_{z}\hat{\mathbf{z}},
\label{eq:HZ}
\end{equation}
with
\begin{equation}
\hat{H}_{\epsilon} := \hat{g}^{\epsilon}\mu_{B}B_{\epsilon} = \hat{g}^{\epsilon}\mathcal{B}_{\epsilon},
\end{equation}
and where we remind that $\mathcal{B}_{\epsilon} :=\mu_{B}B_{\epsilon}$, with $\epsilon=x,y,z$ and $\mu_B$ is the Bohr magneton.

The components of the magnetic field Hamiltonian transform as pseudo-vectors, \textit{i.e.}:
\begin{align}
\left\{ H_{x},H_{y}\right\}  & \sim\left\{ R_{x},R_{y}\right\} \sim\Gamma_{5}\nonumber \\
H_{z} & \sim R_{z}\sim\Gamma_{2} \, .
\end{align}

By performing the direct product between the irreps of the excitons, we can immediately reveal how the external magnetic field will couple the different exciton states. We find
\begin{align}
A-A:\;\Gamma_{6}\otimes\Gamma_{6} & =\Gamma_{1}\oplus\Gamma_{2}\oplus\Gamma_{6},\nonumber \\
A-G:\;\Gamma_{6}\otimes\Gamma_{4} & =\Gamma_{5},\nonumber \\
A-D:\;\Gamma_{6}\otimes\Gamma_{3} & =\Gamma_{5},\nonumber \\
G-D:\;\Gamma_{4}\otimes\Gamma_{3} & =\Gamma_{2}, 
\end{align}
which indicates that the A exciton subspace couples with $\mathbf{B} \parallel \hat{\mathbf{z}}$, that G and D excitons couple with $\mathbf{B} \parallel \hat{\mathbf{z}}$, and that the $A$ exciton couples with $G$ and $D$ excitons via $\mathbf{B} \perp \hat{\mathbf{z}}$.

Using this information, the Hamiltonian in Eq. \eqref{eq:HZ} reads in matrix form
\begin{equation}
H_{\text{Z}}(\mathbf{B})=\left[\begin{array}{ccc}
g_{A}\mathcal{B}_{z}\otimes M_{A} & g_{AG}\mathcal{B}_{xy}\otimes M_{AG} & g_{AD}\mathcal{B}_{xy}\otimes M_{AD}\\
 & 0 & g_{GD}\mathcal{B}_{z} \otimes M_{GD}\\
\text{c.c.} &  & 0
\end{array}\right],
\label{eq:HZsym1}
\end{equation}
where $g_{A}$ is the A-exciton g-factor, $g_{GD}$ is the g-factor coupling G and D excitons, $g_{AG}$ and $g_{BD}$ are g-factors that coupling A-G and A-D excitons, respectively. The matrices $M_{A}$ ($2\times2$), $M_{AG}$ ($2\times1$), $M_{AD}$ ($2\times1$), and $M_{GD}$ ($1\times1$) incorporate complex values related to the particular choice of the exciton basis. We emphasize that the most crucial information given by the symmetry-based Hamiltonian is that $\mathbf{B} \parallel \hat{\mathbf{z}}$ cannot mix $A$ and $G,D$ excitons. This can only be achieved with in-plane ($\mathbf{B} \perp \hat{\mathbf{z}}$) fields. 

Let us further elaborate on the choice of the basis set to the Hamiltonian \ref{eq:HZsym1}. The symmetry relations for pseudovector operators acting on the states of Eq.~\ref{eq:exc_irreps} are given by
\begin{align}
\left\langle \Gamma_{6}^{1}\left|w_{z}\right|\Gamma_{6}^{2}\right\rangle  & =-\left\langle \Gamma_{6}^{2}\left|w_{z}\right|\Gamma_{6}^{1}\right\rangle, \nonumber \\
\left\langle \Gamma_{6}^{1}\left|w_{x}\right|\Gamma_{3}\right\rangle  & =\left\langle \Gamma_{6}^{2}\left|w_{y}\right|\Gamma_{3}\right\rangle, \nonumber \\
\left\langle \Gamma_{6}^{2}\left|w_{x}\right|\Gamma_{4}\right\rangle  & =-\left\langle \Gamma_{6}^{1}\left|w_{y}\right|\Gamma_{4}\right\rangle. 
\end{align}

By choosing the basis set as $\left\{ \left|\Gamma^{+}_{6}\right\rangle, \left|\Gamma^{-}_{6}\right\rangle, \left|\Gamma_{4}\right\rangle, \left|\Gamma_{3}\right\rangle \right\}$ ($\Gamma^{\pm}_{6}$ relates to fully circularly polarization, s$_\pm$), the Hamiltonian takes the form
\begin{equation}
H_{\text{Z}}(\mathbf{B}) = \left[\begin{array}{cccc}
+g_{A}\mathcal{B}_{z} & 0 & i\alpha_{AG}g_{AG}\mathcal{B}_{-} & \alpha_{AD}g_{AD}\mathcal{B}_{-}\\
0 & -g_{A}\mathcal{B}_{z} & -i\alpha_{AG}g_{AG}\mathcal{B}_{+} & \alpha_{AD}g_{AD}\mathcal{B}_{+}\\
 &  & 0 & \alpha_{GD}g_{GD}\mathcal{B}_{z}\\
\text{c.c} &  & \alpha_{GD}^{*}g_{GD}\mathcal{B}_{z} & 0
\end{array}\right],
\label{eq:HZsym2}
\end{equation}
with $\mathcal{B}_{\pm}=\left(\mathcal{B}_{x}\pm i\mathcal{B}_{y}\right) / \sqrt{2}$. Note how the coefficients depend on the choice of basis. In the numerical calculations within the GW-BSE formalism, we do not expect the Hamiltonian to take such a intuitive form since the exciton basis will very likely a superposition of this well-defined basis. Such ``mixed" basis are not only restricted to  GW-BSE calculations. They are, in fact, a common outcome from first principles calculations whenever the bands (or irreps) are degenerate (or nearly degenerate)\cite{Kurpas2016PRB, Kurpas2019PRB, Cassiano2024SciPost, Zhang2023vasp2kp}.

At this point, the non-zero matrix elements allowed by symmetry (Eqs.~\ref{eq:HZsym1} and ~\ref{eq:HZsym2}) provide the right functional form for the magnetic field dependence, which is particularly relevant for fitting experimental data. However, the g-factor terms do not have a precise  meaning without a proper connection to the relevant energies and states that constitute each exciton. In order to do so and provide a microscopic foundation for the symmetry-based Hamiltonian, we can approximate the exciton states by the coefficient directly at the K valleys, as previously considered in other works\cite{Robert2017PRB, Molas2017TDM, Molas2019PRL}
\begin{align}
\left|A_{+}\right\rangle & = \left|\text{VB}_{+},K\right\rangle \left|\text{CB}_{+},K\right\rangle, \nonumber \\
\left|A_{-}\right\rangle & = \left|\text{VB}_{+},-K\right\rangle \left|\text{CB}_{+},-K\right\rangle, \nonumber \\
\left|G\right\rangle & =\frac{1}{\sqrt{2}}\left( \left|\text{VB}_{+},K\right\rangle \left|\text{CB}_{-},K\right\rangle + \left|\text{VB}_{+},-K\right\rangle \left|\text{CB}_{-},-K\right\rangle \right),\nonumber \\
\left|D\right\rangle & =\frac{1}{\sqrt{2}}\left( \left|\text{VB}_{+},K\right\rangle \left|\text{CB}_{-},K\right\rangle - \left|\text{VB}_{+},-K\right\rangle \left|\text{CB}_{-},-K\right\rangle \right).
\label{eq:exc_basisK}
\end{align}

To verify that this approximation to the full exciton states  satisfies the symmetry considerations of Eq.~\eqref{eq:exc_irreps}, we can evaluate the optical selection rules given by the matrix elements
\begin{equation}
\left\langle \emptyset \left|\mathbf{p}\right |S \right\rangle =\sum_{vc\mathbf{k}}\mathcal{A}_{vc\mathbf{k}}^{S}\left\langle c\mathbf{k}\left|\mathbf{p}\right|v\mathbf{k}\right\rangle, 
\end{equation}
where, $|\emptyset \rangle$ corresponds to the ground state (Fermi sea), $S\in\{A_{\pm},G,D\}$ and $\mathcal{A}_{vc\mathbf{k}}^{S}$ is the
exciton amplitude of the $S$-th exciton, solution of the
BSE, see Eq. \eqref{e0}.

The single-particle selection rules are given by
\begin{align}
\left\langle \text{CB}_{+},\pm K\left|\mathbf{p}\cdot\frac{\left(\hat{x}\pm i\hat{y}\right)}{\sqrt{2}}\right|\text{VB}_{+},\pm K\right\rangle  & =\gamma\left(\frac{1\pm i}{\sqrt{2}}\right),\nonumber \\
\\ \nonumber
\left\langle \text{CB}_{-},+K\left|\mathbf{p}\cdot\hat{z}\right|\text{VB}_{+},+K\right\rangle  & =\left|\left\langle \text{CB}_{-},-K\left|\mathbf{p}\cdot\hat{z}\right|\text{VB}_{+},-K\right\rangle \right| = \gamma_{z},
\end{align}
thus, it follows that the exciton selection rules are given by
\begin{align}
\left|\left\langle A_{\pm}\left|\mathbf{p}\cdot\hat{\alpha}\right|0\right\rangle \right|^{2} & =\left|\left\langle \text{CB}_{+},\pm K\left|\mathbf{p}\cdot\left(\frac{\hat{x}\pm i\hat{y}}{\sqrt{2}}\right)\right|\text{VB}_{+},\pm K\right\rangle \right|^{2},\nonumber \\
 & =\gamma^{2}.
\end{align}
\begin{align}
\left|\left\langle G\left|\mathbf{p}\cdot\hat{\alpha}\right|0\right\rangle \right|^{2} & =\frac{1}{2}\left|\left\langle \text{CB}_{-},K\left|\mathbf{p}\cdot\hat{z}\right|\text{VB}_{+},K\right\rangle +\left\langle \text{CB}_{-},-K\left|\mathbf{p}\cdot\hat{z}\right|\text{VB}_{+},-K\right\rangle \right|^{2},\nonumber \\
 & =\frac{1}{2}\left|\gamma_{z}+\gamma_{z}\right|^{2},\nonumber \\
 & =2\gamma_{z}^{2}.
\end{align}
\begin{align}
\left|\left\langle D\left|\mathbf{p}\cdot\hat{\alpha}\right|0\right\rangle \right|^{2} & =\frac{1}{2}\left|\left\langle \text{CB}_{-},K\left|\mathbf{p}\cdot\hat{z}\right|\text{VB}_{+},K\right\rangle -\left\langle \text{CB}_{-},-K\left|\mathbf{p}\cdot\hat{z}\right|\text{VB}_{+},-K\right\rangle \right|^{2},\nonumber \\
 & =\frac{1}{2}\left|\gamma_{z}-\gamma_{z}\right|^{2},\nonumber \\
 & =0,
\end{align}
with $\gamma$ and $\gamma_z$ differing by 2 orders of magnitude ($\gamma \gg \gamma_z$). It is important to emphasize here that $\gamma_z$ is rather small because of the small overlap between the wavefunctions with nearly opposite spin, i. e., $\text{VB}_{+}$ (nearly totally spin-up at $K$) and $\text{CB}_{-}$ (nearly totally spin-down at $K$)\cite{FariaJunior2022NJP}.

Besides evaluating the dipole matrix elements for these excitonic transitions, we also verify that this basis set satisfy the irreps of Eq.~\ref{eq:exc_irreps} with respect to the mirror plane symmetry operation, $\hat{\sigma_v}$ (crossing the M points of the first Brillouin zone, as shown in Fig.~1 of the main text). By inspecting the character table of the $D_{3h}$ group for the mirror plane(s) symmetry operation, we extract a character of $0$ for the $A$ excitons, $1$ for the $G$ exciton, $-1$ for the $D$ exciton. The effect of $\hat{\sigma_v}$ on the exciton basis can be written as
\begin{align}
\hat{\sigma_v} A_+ & = A_-, \\
\hat{\sigma_v} A_- & = A_+, \\
\hat{\sigma_v} G & = G, \\
\hat{\sigma_v} D & = -D,
\end{align}
in which the sign for the $D$ exciton is directly connected to the sign change between $K$ and $-K$ points, as given in Eq.~\ref{eq:exc_basisK}, consequence of intervalley exchange coupling \cite{Molas2019PRL}.

Let us now evaluate the exciton spin matrix elements for magnetic fields along $z$ and along $x$ directions using Eqs.~\eqref{eq:spin_bloch} and \eqref{e1}. For the subset $\left|\text{CB}_{+},\pm K\right\rangle,\left|\text{CB}_{-},\pm K\right\rangle,\left|\text{VB}_{+},\pm K\right\rangle$,
numerical evaluations within DFT give
\begin{equation}
\Sigma_{K}^{z}=\left[\begin{array}{ccc}
\sigma_{\text{CB}_{+}} & 0 & 0\\
0 & \sigma_{\text{CB}_{-}} & 0\\
0 & 0 & \sigma_{\text{VB}_{+}}
\end{array}\right],
\qquad
\Sigma_{K}^{x}=\left[\begin{array}{ccc}
0 & \sigma_{\text{CB}} & 0\\
\sigma_{\text{CB}} & 0 & 0\\
0 & 0 & 0
\end{array}\right],
\qquad
\Sigma_{-K}^{\epsilon}=-\Sigma_{K}^{\epsilon}
\end{equation}
with the explicit matrix elements expressed as
\begin{align}
\sigma_{\text{CB}_{+}} & = \left\langle {\text{CB}_{+}} \left| \sigma_z \right| {\text{CB}_{+}} \right\rangle, \nonumber \\ 
\sigma_{\text{CB}_{-}} & = \left\langle {\text{CB}_{-}} \left| \sigma_z \right| {\text{CB}_{-}} \right\rangle, \nonumber \\
\sigma_{\text{VB}_{+}} & = \left\langle {\text{VB}_{+}} \left| \sigma_z \right| {\text{VB}_{+}} \right\rangle, \nonumber \\
\sigma_{\text{CB}} & = \left\langle {\text{CB}_{+}} \left| \sigma_{x(y)} \right| {\text{CB}_{-}} \right\rangle.
\end{align}
The values obtained using \textsc{Quantum Espresso} and Wien2k are given in Table~\ref{tab:DFTslg}.

In this basis, the exciton spin matrix elements in the $\epsilon = z$ direction are
\begin{align}
\langle S|\hat{\Sigma}^{z}|S'\rangle & =\left[(\mathcal{A}_{\text{VB}_{+},\text{CB}_{+},+K}^{S})^{\ast}\mathcal{A}_{\text{VB}_{+},\text{CB}_{+},+K}^{S'}-(\mathcal{A}_{c\text{VB}_{+},\text{CB}_{+},-K}^{S})^{\ast}\mathcal{A}_{\text{VB}_{+},\text{CB}_{+},-K}^{S'}\right]\left[\sigma_{\text{CB}_{+}}-\sigma_{\text{VB}_{+}}\right] \nonumber \\
 & +\left[(\mathcal{A}_{\text{VB}_{+},\text{CB}_{-},+K}^{S})^{\ast}\mathcal{A}_{\text{VB}_{+},\text{CB}_{-},+K}^{S'}-(\mathcal{A}_{\text{VB}_{+},\text{CB}_{-},-K}^{S})^{\ast}\mathcal{A}_{\text{VB}_{+},\text{CB}_{-},-K}^{S'}\right]\left[\sigma_{\text{CB}_{-}}-\sigma_{\text{VB}_{+}}\right],
\end{align}

\begin{align}
S=S'\Rightarrow\langle S|\hat{\Sigma}^{\epsilon}|S\rangle & =\left[\left|\mathcal{A}_{\text{VB}_{+},\text{CB}_{+},+K}^{S}\right|^{2}-\left|\mathcal{A}_{\text{VB}_{+},\text{CB}_{+},-K}^{S}\right|^{2}\right]\left[\sigma_{\text{CB}_{+}}-\sigma_{\text{VB}_{+}}\right]\nonumber \\
 & +\left[\left|\mathcal{A}_{\text{VB}_{+},\text{CB}_{-},+K}^{S}\right|^{2}-\left|\mathcal{A}_{\text{VB}_{+},\text{CB}_{-},-K}^{S}\right|^{2}\right]\left[\sigma_{\text{CB}_{-}}-\sigma_{\text{VB}_{+}}\right],
\end{align}

\begin{equation}
\hat{\Sigma}^{z}=\left[\begin{array}{cc|cc}
+\sigma_{\text{CB}_{+}}-\sigma_{\text{VB}_{+}} & 0 & 0 & 0\\
0 & -\sigma_{\text{CB}_{+}}+\sigma_{\text{VB}_{+}} & 0 & 0\\
\hline 0 & 0 & 0 & \sigma_{\text{CB}_{-}}-\sigma_{\text{VB}_{+}}\\
0 & 0 & \sigma_{\text{CB}_{-}}-\sigma_{\text{VB}_{+}} & 0
\end{array}\right].
\end{equation}

The exciton spin matrix elements for $\epsilon = x$ are
\begin{align}
\langle S|\hat{\Sigma}^{x}|S'\rangle & =\Bigg[(\mathcal{A}_{\text{CB}_{+},\text{VB}_{+},+K}^{S})^{\ast}\mathcal{A}_{\text{CB}_{-},\text{VB}_{+},+K}^{S'}-(\mathcal{A}_{\text{CB}_{+},\text{VB}_{+},-K}^{S})^{\ast}\mathcal{A}_{\text{CB}_{-},\text{CB}_{+},-K}^{S'}\Bigg]\sigma_{\text{CB}}\nonumber \\
 & +\Bigg[(\mathcal{A}_{\text{CB}_{-},\text{CB}_{+},+K}^{S})^{\ast}\mathcal{A}_{\text{CB}_{+},\text{VB}_{+},+K}^{S'}-(\mathcal{A}_{\text{CB}_{-},\text{VB}_{+},-K}^{S})^{\ast}\mathcal{A}_{\text{CB}_{+},\text{CB}_{+},-K}^{S'}\Bigg]\sigma_{\text{CB}},
\end{align}

\begin{equation}
\hat{\Sigma}^{x}=\frac{\sigma_{\text{CB}}}{\sqrt{2}}\left[\begin{array}{cc|cc}
0 & 0 & 1 & 1\\
0 & 0 & -1 & 1\\
\hline 1 & -1 & 0 & 0\\
1 & 1 & 0 & 0
\end{array}\right].
\end{equation}

We finally obtain the total Hamiltonian, $\hat{H} = \hat{H}^\textnormal{BSE} + \hat{\mathbf{g}} \cdot \bm{\mathcal{B}}$, including the zero field diagonal part and a Zeeman term for a magnetic field along the the $xz$ plane
\begin{equation}
H=\left[\begin{array}{cc|cc}
\Delta+g_{A}\mathcal{B}\cos\theta & 0 & g_{AG}\mathcal{B}\sin\theta & g_{AD}\mathcal{B}\sin\theta\\
0 & \Delta-g_{\text{A}}\mathcal{B}\cos\theta & -g_{AG}\mathcal{B}\sin\theta & g_{AD}\mathcal{B}\sin\theta\\
\hline g_{\text{AG}}\mathcal{B}\sin\theta & -g_{AG}\mathcal{B}\sin\theta & \delta & g_{GD}\mathcal{B}\cos\theta\\
g_{AD}\mathcal{B}\sin\theta & g_{AD}\mathcal{B}\sin\theta & g_{GD}\mathcal{B}\cos\theta & 0
\end{array}\right].
\end{equation}
\\
with $\mathcal{B}=\mu_{B}B$. We have two limiting cases, $\theta=0\Rightarrow\mathbf{B}=B\hat{\mathbf{z}}$ and $\theta=90^{\circ}\Rightarrow\mathbf{B}=B\hat{\mathbf{x}}$. The exciton energy splittings can be fitted to our GW-BSE results, $\Delta = 54.8 \; \textrm{meV}$ and $\delta = 2.4 \; \textrm{meV}$, and the  g-factor values can be numerically calculated directly at the K-point, $g_{A} = g_{\text{CB}_+} - g_{\text{VB}_+} = -2.1$, $g_{GD} = g_{\text{CB}_-} - g_{\text{VB}_+} = -4.371$, $g_{AG} = g_{AD} = g_{\text{CB}}/\sqrt{2} = 0.847$. The (single-particle) band g-factors are obtained from the \textsc{Quantum Espresso} calculations with a GW scissor shift operation\cite{Amit2022}, and they are given in Table~\ref{tab:DFTslg}. In the same table, we also compare the \textsc{Quantum Espresso} calculated values with WIEN2k to verify that both DFT codes give very similar results.

\setlength{\tabcolsep}{10pt}
\begin{table}[H]
\caption{Numerical values for spin and orbital angular momenta. In parentheses, DFT indicates that the operator was calculated using DFT energies, GW refers to a rigid scissor shift correction using the GW band gap obtained at the K point, and SC denotes a rigid scissor shift of 1 eV. While the values of $\sigma$ and $L$ in the $z$-direction are real within the calculations, they are complex for the $x, y$ directions due to the presence of (arbitrary) complex phases. However, for our purposes here, it suffices to consider the absolute value.}
\begin{center}
\begin{tabular}{cccrcc}
\hline
\hline
 &  & $\ensuremath{\text{VB}_{+}}$ & $\ensuremath{\text{CB}_{-}}$ & $\ensuremath{\text{CB}_{+}}$ & $\ensuremath{\text{CB}}(xy)$\tabularnewline
\hline 
QUANTUM  & $\sigma$  & 0.999 & $-$0.917 & 0.977 & 0.973\tabularnewline
ESPRESSO & $L$ (DFT) & 5.348 &    2.040 & 3.134 & 0.209\tabularnewline
         & $g$ (DFT) & 6.347 &    1.123 & 4.111 & 1.182\tabularnewline
         & $L$ (GW)  & 3.754 &    1.199 & 1.676 & 0.225\tabularnewline
         & $g$ (GW)  & 4.753 &    0.282 & 2.653 & 1.198\tabularnewline
\hline 
WIEN2k & $\sigma$  & 0.999 & $-$0.921 & 0.977 & 0.974 \tabularnewline
       & $L$ (DFT) & 5.310 &    1.962 & 3.140 & 0.064 \tabularnewline
       & $g$ (DFT) & 6.310 &    1.042 & 4.117 & 1.038 \tabularnewline
       & $L$ (SC)  & 3.716 &    1.155 & 1.676 & 0.088 \tabularnewline
       & $g$ (SC)  & 4.716 &    0.234 & 2.656 & 1.062\tabularnewline
 \hline
 \hline
\end{tabular}
\end{center}
\label{tab:DFTslg}
\end{table}


\clearpage

\newpage

\section{Additional computational data}

\subsection{GW-BSE results of $S\in\{A_{\pm},G,D\}$}

Table \ref{tab:monoBSE} we summarize the energies and oscillator strengths calculated with the GW-BSE formalism.

\begin{table}[H]
\caption{Excitonic energies and oscillator strengths for the low-energy $\{S\} = \{A_{\pm},G,D\}$ in monolayer WSe\textsubscript{2}. Energy values are given in meV and oscillator strengths in units of $e^2 a_0^2$. The (bright-grey) A-G splitting is 52.4 meV while the (grey-dark) G-D splitting is 2.4 meV. This subspace is well-separated of the other excitons, as the next exciton state is $\sim 164$ meV above the $A$ exciton.}
\begin{center}
\begin{tabular}{cccccc}
\hline
\hline
Exciton & Energy & $\left|P_{+}\right|^{2}$ & $\left|P_{-}\right|^{2}$ & $\left|P_{z}\right|^{2}$  & Irrep \\
\hline
 \multirow{2}{*}{A} & $\;$1771.58$\;$ & $\;$783.025 & $\;$4771.892$\;$ & -- &  \multirow{2}{*}{$\Gamma_{6}$} \\
    & 1771.52 & $\;$4775.531$\;$ & $\;$786.309 & -- & \\
G & 1719.14 & -- & -- & 9.562 & $\Gamma_{4}$ \\
D & 1716.77 & -- & -- & 0.001 & $\Gamma_{3}$ \\
\hline
\hline
\end{tabular}
\end{center}
\label{tab:monoBSE}
\end{table}

The numerical values of the spin matrices obtained via Eq.~\ref{e1} and the orbital angular momentum matrices obtained via Eq.~\ref{eq:L_exc} in the basis of $S\in\{A_{\pm},G,D\}$ are shown below. Despite the numerical hybridization of exciton states in the degenerate subspaces, the spin and orbital matrices in the exciton basis fully satisfy the symmetry considerations, clearly visible in the blocks with zero elements (between A and G/D for $\mathbf{B} \parallel z$ and within the A and G/D blocks for $\mathbf{B} \parallel x,y$).

\lstset{language=code}
\begin{lstlisting}
Sigma-x = 
[ 0.00+0.00j  0.00+0.00j  0.07+0.14j  0.03+0.07j]
[ 0.00+0.00j  0.00+0.00j -0.12+0.12j -0.03+0.04j]
[ 0.07-0.14j -0.12-0.12j  0.00+0.00j  0.00+0.00j]
[ 0.03-0.07j -0.03-0.04j  0.00+0.00j  0.00+0.00j]

Sigma-y =
[ 0.00+0.00j  0.00+0.00j -0.15+0.07j -0.04+0.03j]
[-0.00+0.00j  0.00+0.00j -0.11-0.10j -0.05-0.06j]
[-0.15-0.07j -0.11+0.10j  0.00+0.00j  0.00+0.00j]
[-0.04-0.03j -0.05+0.06j  0.00+0.00j -0.00+0.00j]


Sigma-z =
[-0.04+0.00j -0.62+1.81j  0.00+0.00j -0.00+0.00j]
[-0.62-1.81j  0.04+0.00j -0.00+0.00j  0.00+0.00j]
[-0.00+0.00j  0.00+0.00j -0.02+0.00j -0.02-0.00j]
[ 0.00+0.00j -0.00+0.00j -0.02+0.00j  0.02+0.00j]
\end{lstlisting}

\lstset{language=code}
\begin{lstlisting}
L-x =
[ 0.00+0.00j  0.00+0.00j -0.01-0.03j -0.01-0.02j]
[ 0.00+0.00j  0.00+0.00j  0.03-0.02j  0.01-0.01j]
[-0.01+0.03j  0.03+0.02j  0.00+0.00j  0.00+0.00j]
[-0.01+0.02j  0.01+0.01j  0.00+0.00j  0.00+0.00j]

L-y =
[ 0.00+0.00j  0.00+0.00j  0.03-0.02j  0.01-0.01j]
[-0.00-0.00j  0.00+0.00j  0.02+0.02j  0.01+0.01j]
[ 0.03+0.02j  0.02-0.02j  0.00+0.00j -0.00+0.00j]
[ 0.01+0.01j  0.01-0.01j  0.00+0.00j  0.00+0.00j]

L-z =
[-0.04+0.00j -0.65+1.90j  0.00+0.00j  0.00+0.00j]
[-0.65-1.90j  0.05+0.00j  0.00+0.00j -0.00+0.00j]
[ 0.00+0.00j -0.00+0.00j -1.24+0.00j -1.20-0.01j]
[ 0.00+0.00j  0.00+0.00j -1.20+0.01j  1.24+0.00j]
\end{lstlisting}


\clearpage

\newpage

\section{Additional figures}

\begin{figure}[H]
\begin{center} 
\includegraphics[scale=0.95]{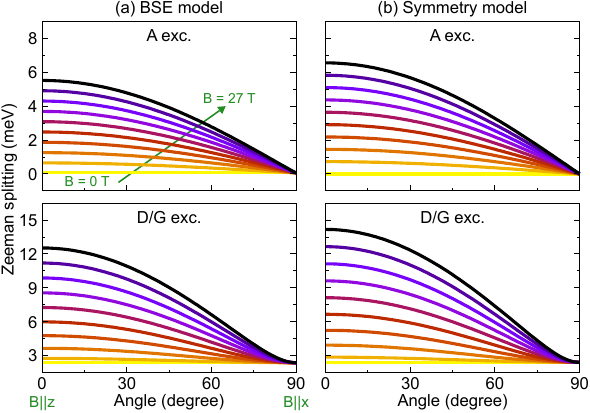}
\caption{Zeeman splitting of the A and D/G excitons as a function of the angle for different values of magnetic field for (a) the BSE model and (b) the symmetry-adapted model.}
\end{center}
\end{figure}

\begin{figure}[H]
\begin{center} 
\includegraphics[width=0.95\textwidth]{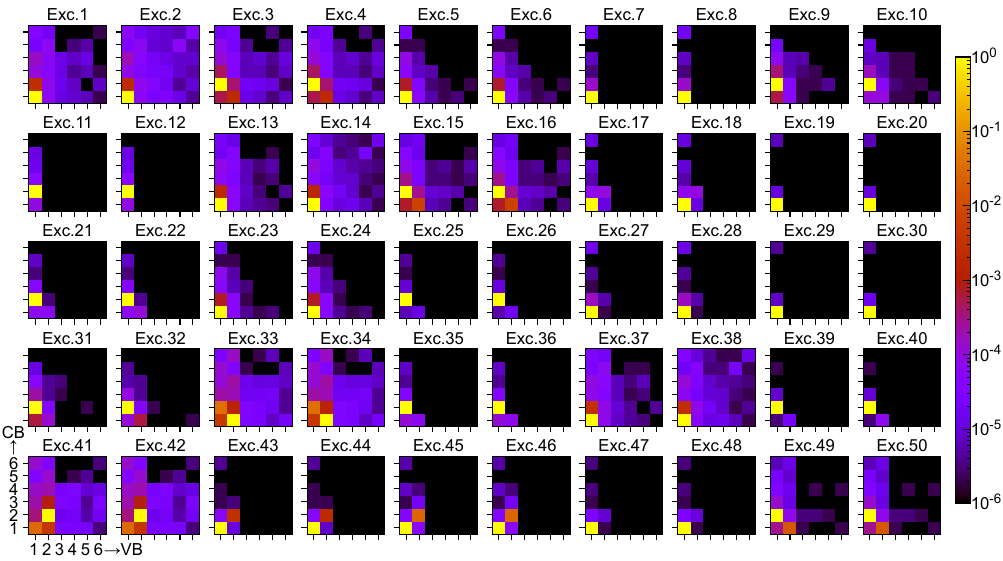}
\caption{Band composition of the exciton wavefunctions.}
\end{center}
\end{figure}

\begin{figure}[H]
\begin{center} 
\includegraphics[width=0.95\textwidth]{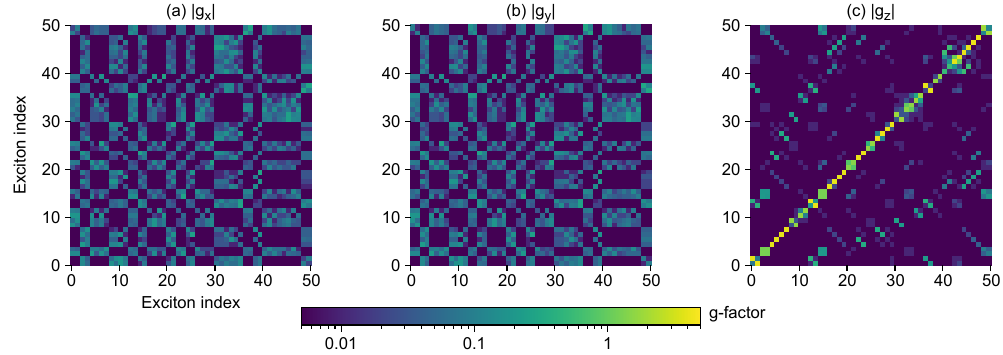}
\caption{Exciton g-factor matrix}
\end{center}
\end{figure}

\begin{figure}[H]
\begin{center} 
\includegraphics[width=0.95\textwidth]{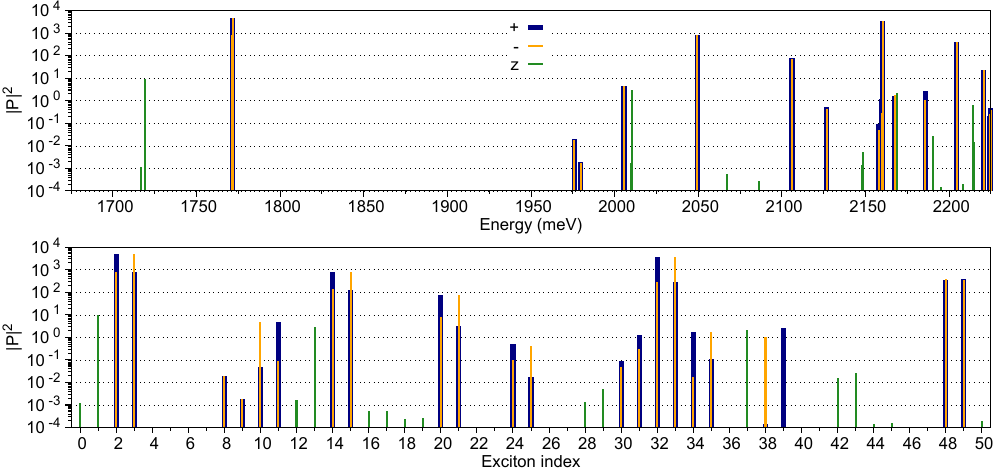}
\caption{Oscillator strength for various excitonic states.}
\end{center}
\end{figure}

\begin{figure}[H]
\begin{center} 
\includegraphics[width=0.95\textwidth]{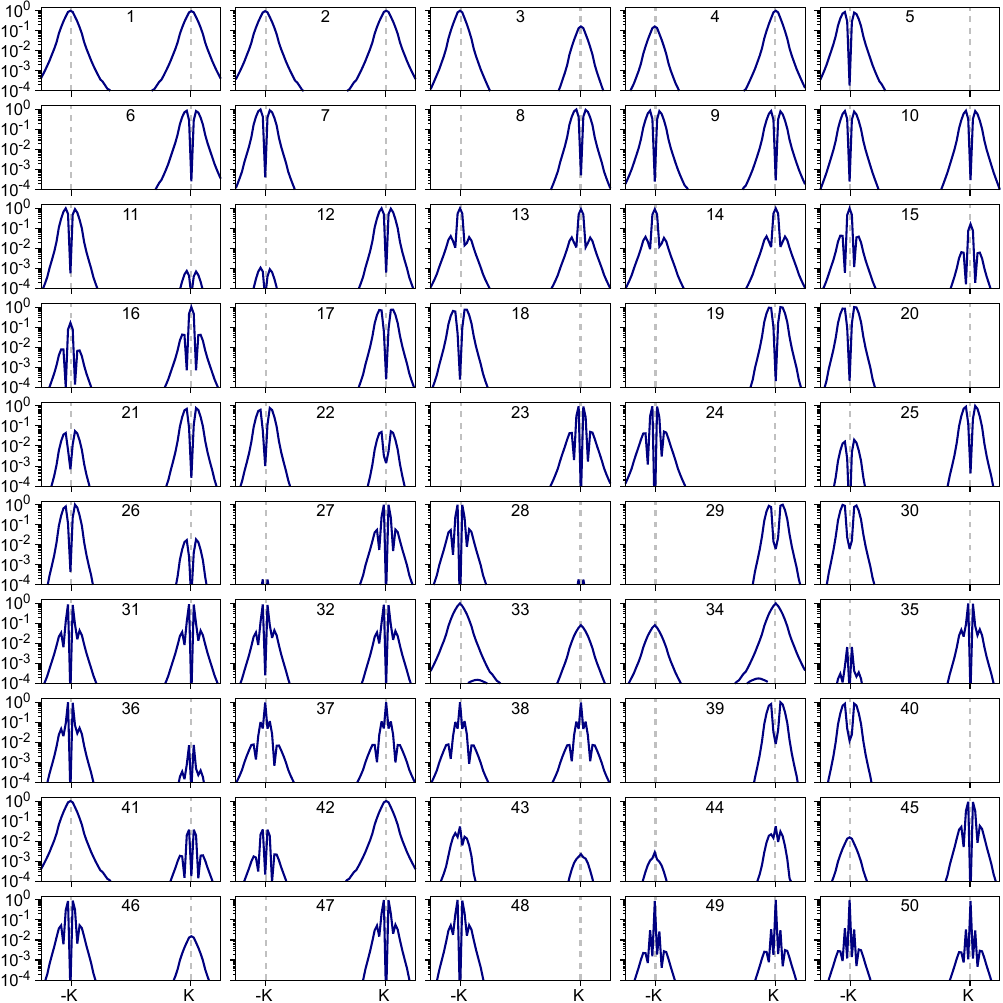}
\caption{Density of the exciton wavefunctions.}
\end{center}
\end{figure}

\begin{figure}[H]
\begin{center} 
\includegraphics[width=0.95\textwidth]{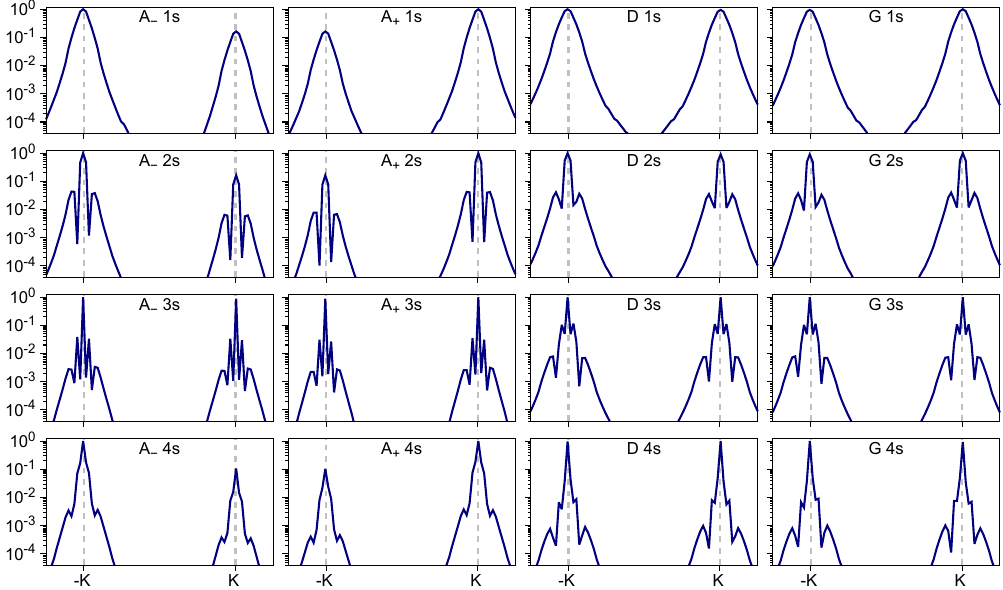}
\caption{Density of the A, D, and G 1s--4s exciton wavefunctions.}
\end{center}
\end{figure}


\bibliography{biblio}
